\documentclass[twocolumn,aps,prl,preprintnumbers]{revtex4}
\usepackage{bbm}
\usepackage[latin9]{inputenc}
\usepackage{amsmath}
\usepackage{amssymb}
\usepackage{graphicx}
\usepackage{mathrsfs}
\usepackage{amsfonts}
\usepackage{amsthm}
\usepackage{color}
\usepackage{txfonts}
\usepackage[colorlinks=true,citecolor=blue,linkcolor=blue,urlcolor=blue,anchorcolor=blue]{hyperref}%
\hypersetup{colorlinks=true,citecolor=blue,linkcolor=blue,urlcolor=blue}
\providecommand{\U}[1]{\protect\rule{.1in}{.1in}}
\makeatletter
\@ifundefined{textcolor}{}
{
\definecolor{BLACK}{gray}{0}
\definecolor{WHITE}{gray}{1}
\definecolor{RED}{rgb}{1,0,0}
\definecolor{GREEN}{rgb}{0,1,0}
\definecolor{BLUE}{rgb}{0,0,1}
\definecolor{CYAN}{cmyk}{1,0,0,0}
\definecolor{MAGENTA}{cmyk}{0,1,0,0}
\definecolor{YELLOW}{cmyk}{0,0,1,0}
}
\makeatother
\begin{document}
\title{Antiferromagnetism emerging in a ferromagnet with gain}
\author{Huanhuan Yang}
\author{C. Wang}
\author{Tianlin Yu}
\author{Yunshan Cao}
\email[]{yunshan.cao@uestc.edu.cn}
\author{Peng Yan}
\email[]{yan@uestc.edu.cn}
\affiliation{School of Electronic Science and Engineering and State Key Laboratory of Electronic Thin Film and Integrated Devices, University of
Electronic Science and Technology of China, Chengdu 610054, China}

\begin{abstract}
We present a theoretical mapping to show that a ferromagnet with gain (loss) is equivalent to an antiferromagnet with an equal amount of loss (gain). Our finding indicates a novel first-order ferromagnet-antiferromagnet phase transition by tuning the gain-loss parameter. As an appealing application, we demonstrate the realization as well as the manipulation of the antiferromagnetic skyrmion, a stable topological quasiparticle not yet observed experimentally, in a chiral ferromagnetic thin film with gain. We also consider ferromagnetic bilayers with balanced gain and loss, and show that the antiferromagnetic skyrmion can be found only in the cases with broken parity-time symmetry phase. Our results pave a way for investigating the emerging antiferromagnetic spintronics and parity-time symmetric magnonics in ferromagnets.
\end{abstract}

\maketitle
The first-order antiferromagnetic (AFM) to ferromagnetic (FM) phase transition (or the other way around) has received tremendous attention in the community of condensed matter physics \cite{Kittel1960,Zakharov1967,Brabers1999,Marti2014,Gatel2017}. It involves a transition from a configuration with an antiparallel orientation of the magnetic moments to a parallel configuration, or vice versa. Conventionally, the FM-AFM phase transition is induced by heating \cite{Stamm2008,Chirkova2017}, pressure \cite{Ricodeau1972,Li2017}, and field \cite{Brabers1999,Marti2014,Lee2015NC}, restricted to some specific materials, such as FeRh, $R$Mn$_{2}$Ge$_{2}$, and Dy, to name a few. It should be very interesting and important if one can find
other effective control methods and principles to manipulate the first-order FM-AFM phase transition without the mentioned constraints.

\begin{figure}[htbp]
	\centering
	\includegraphics[width=0.47\textwidth]{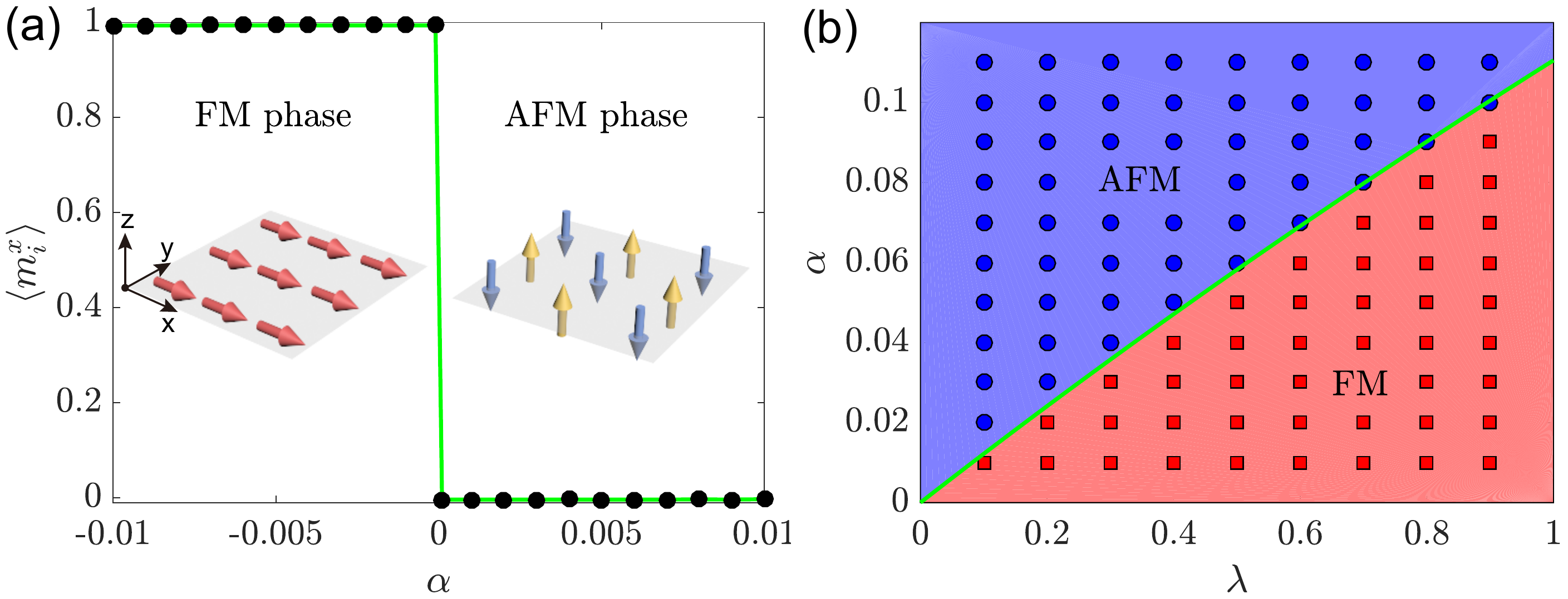}
	\caption{(a) Order parameter $\langle m^x_i\rangle$
	as a function of $\alpha$ in a single FM layer. Dots
	are numerical results.
	Inset: Spin configurations of FM and AFM states. (b) Phase diagram of the $\mathcal{P}\mathcal{T}$
	symmetric bilayer by tuning the gain-loss parameter
	$\alpha$ and the interlayer coupling constant $\lambda$.
	Symbols are numerical results and the green curve
	is the analytical formula
	$\min_{\forall\mathbf{k}}\alpha_{\text{c}}(\mathbf{k})=\lambda/\big[2\sqrt{\zeta_{0}(\lambda+\zeta_{0})}\big]$.}
	\label{fig1}
\end{figure}

Loss and gain are ubiquitous in nature. Tantalizing physics under their balance has attracted enormous interest and found many great applications in the context of parity-time ($\mathcal {P}\mathcal {T}$) symmetry and exceptional point (EP) \cite{Konotop2016} in a broad field of quantum mechanics \cite{Bender1998}, optics \cite{Makris2008,Ruter2010,Regensburger2012,Zhang2018}, acoustics \cite{Zhu2014,Fleury2015}, optomechanics \cite{Lu2015,Xu2016}, electronics \cite{Schindler2011,Schindler2012,Bender2013,Sid2017,Chen2018}, and very recently in spintronics \cite{Lee2015,Yan2015,Galda2016,Galda2017,Galda2018} and cavity spintronics \cite{Harder2017,Zhang2017}. In Ref. \cite{Lee2015}, Lee, Kottos and Shapiro proposed two coupled macroscopic FM layers respecting the $\mathcal {P}\mathcal {T}$ symmetry: one layer with loss and another one with an equal amount of gain, and discussed their dynamics in the framework of Landau-Lifshitz-Gilbert (LLG) equation \cite{Gilbert2004}. The positive Gilbert damping (loss) in magnets usually comes from the phonon dissipation and the electromagnetic radiation, while the negative one (gain) can be realized by parametric driving and/or spin transfer torque \cite{Lee2015,Galda2016,Galda2017,Galda2018}. In this work we investigate the properties of microscopic easy-plane ``gain" ferromagnets. We map the equation of motion of local magnetic moments to a dissipative one in antiferromagnets, and thus argue their equivalence. Based on this finding, we numerically demonstrate the formation of an AFM skyrmion, a stable topological quasiparticle yet to be observed experimentally, in single-layer chiral ferromagnets with gain, and study its dynamics driven by spin-polarized electric currents. We also investigate the spin-wave spectrum in $\mathcal {P}\mathcal {T}$ symmetric bilayer ferromagnets by tuning the balanced gain-loss parameter. The phase diagram of the first-order FM-AFM phase transition is obtained (see Fig.~\ref{fig1}). It is interesting that the emerging antiferromagnetism and AFM skyrmion can only be found when the $\mathcal {P}\mathcal {T}$ symmetry is broken.

We start with the following Hamiltonian
of a ferromagnet in two spatial dimensions (the $xy$ plane)
\begin{equation}
\begin{aligned}
\mathcal{H}\Big\{\mathbf{m}_i\Big\}=&-\sum_{\langle ij\rangle}
J\mathbf{m}_i \cdot \mathbf{m}_j-\sum_{\langle ij\rangle}
\mathbf{D}_{ij}\cdot(\mathbf{m}_i\times
\mathbf{m}_j)+\sum_{i}K(\mathbf{m}_{i}\cdot\hat{z})^{2}\\
&+\mathcal{H}_{\text{DDI}}\Big\{\mathbf{m}_i\Big\},
\end{aligned}\label{Hamiltonian1}
\end{equation}
where $\mathbf{m}_i$ is the unit spin vector at the
$i$-th site $(i_{x}a,i_{y}a)$ with $i_{x(y)}$ an
arbitrary integer and $a$ the lattice constant,
$J>0$ is the FM exchange coupling
constant, $\mathbf{D}_{ij}=D\hat{r}_{ij}\times \hat{z}$
is the interfacial Dzyaloshinskii-Moriya interaction (DMI) vector \cite{Dzyaloshinskii1958,Moriya1960} with the unit
vector $\hat{r}_{ij}=\mathbf{r}_{ij}/r_{ij}$
connecting sites $i$ and $j$ at a distance
$r_{ij}=|\mathbf{r}_{ij}|$, $\langle ij\rangle$
sums over all nearest-neighbour sites, $K>0$ is
the easy-plane magnetic anisotropy constant, and
 \begin{equation}\label{DDI}
 \mathcal{H}_{\text{DDI}}\Big\{\mathbf{m}_i\Big\}=\sum_{i\neq j}\frac{\mu_{0}M_{s}^{2}a^{6}}{4\pi r^{3}_{ij}}\Big[\mathbf{m}_{i}\cdot\mathbf{m}_{j}-3(\mathbf{m}_{i}\cdot\hat{r}_{ij})(\mathbf{m}_{j}\cdot\hat{r}_{ij})\Big]
\end{equation}
is the non-local dipolar interaction with $\mu_{0}$
the vacuum magnetic permeability and $M_{s}$ the
saturation magnetization. Ideal magnetic materials for
the energy model \eqref{Hamiltonian1} are
Fe$_{0.7}$Co$_{0.3}$Si \cite{Vousden2016}, CoFeB \cite{Xu2018}, etc. The
generalization of the present model to generic two-dimensional (2D) magnets with, e.g., honeycomb lattices,
next-nearest-neighbor interactions, or different anisotropy axes, is straightforward.
\par

In a ferromagnet with gain, the time evolution of the magnetization
dynamics can be described by the modified LLG equation \cite{Lee2015,Gilbert2004}:
\begin{equation}\label{MLLG}
  \frac{d \mathbf{m}_{i}}{d t}=-\gamma \mathbf{m}_{i}\times \mathbf{H}_{\text{eff},i}-\alpha\mathbf{m}_{i}\times\frac{d \mathbf{m}_{i}}{d t},
\end{equation}
where $\gamma$ is the (positive) gyromagnetic ratio and
$\alpha>0$ is the gain coefficient. The first term in the
right-hand side of Eq.~\eqref{MLLG} describes the Larmor
precession of local spins about the effective field
$\mathbf{H}_{\text{eff},i}=-(\mu_{0}M_{s}a^{3})^{-1}\partial\mathcal{H}\Big\{\mathbf{m}_i\Big\}/\partial\mathbf{m}_{i}$. The second term is a torque driving the spin away
from the field. Due to the very presence of the gain,
the energy change rate of the spin system
\begin{equation}\label{Rate}
  \frac{d\mathcal{H}}{dt}=\frac{\alpha\gamma\mu_{0}M_{s}a^{3}}{1+\alpha^{2}}\sum_{i}\big|\mathbf{m}_{i}\times\mathbf{H}_{\text{eff},i}\big|^{2}
\end{equation}
is always nonnegative. The parallel state
of magnetizations in the ferromagnet is thus
unstable, and the system seeks the energy maximum.
\par

To obtain more insights, we utilize a
mapping $\mathbf{n}_{i}=-\mathbf{m}_{i}$
and recast Eq.~\eqref{MLLG} into
\begin{equation}\label{Recast}
  \frac{d \mathbf{n}_{i}}{d t}=-\gamma \mathbf{n}_{i}\times \widetilde{\mathbf{H}}_{\text{eff},i}+\alpha\mathbf{n}_{i}\times\frac{d \mathbf{n}_{i}}{d t},
\end{equation}
which recovers the dissipative LLG equation
describing the collective motion of spin
vectors $\mathbf{n}_{i}$ governed by a new
Hamiltonian $\widetilde{\mathcal{H}}\Big\{\mathbf{n}_i\Big\}=-\mathcal{H}\Big\{\mathbf{n}_i\Big\}$
with $\widetilde{\mathbf{H}}_{\text{eff},i}=-(\mu_{0}M_{s}a^{3})^{-1}\partial\widetilde{\mathcal{H}}\Big\{\mathbf{n}_i\Big\}/\partial\mathbf{n}_{i}$.
Interestingly, we note that $\widetilde{\mathcal{H}}$
can be interpreted as a 2D AFM
Hamiltonian with $J$ the AFM exchange
constant and $K$ the easy-axis anisotropy constant
along $z-$direction. While the physical meaning
of a negative dipolar interaction $-\mathcal{H}_{\text{DDI}}\Big\{\mathbf{n}_i\Big\}$
is not so transparent, we find that the dipole-dipole interaction
effectively renormalizes the exchange constant as $J\rightarrow J-\mu_0 M^2_s a^3/(4\pi)$ when the stabilized magnetizations are aligned in an antiparallel manner, by expanding the Hamiltonian for nearest neighbors. This correction, however, usually is negligibly small: In Fe$_{0.7}$Co$_{0.3}$Si \cite{Vousden2016}, for example, the ratio $\mu_0 M^2_s a^3/(4\pi J)\sim 10^{-4}$. We thus conclude that a ferromagnet with gain is equivalent to an antiferromagnet with an equal amount of loss. The statement can be presented the other way around as well: an antiferromagnet with gain is equivalent to a ferromagnet with the same loss.

Then we show that the above mapping indicates a first-order FM-AFM phase transition by tuning the gain-loss parameter $\alpha$. To this end, we choose the magnetization
$\langle m^x_i \rangle$ as the order parameter, with $\langle\cdots\rangle$ representing the average
over all sites, and numerically
solve Eq.~\eqref{MLLG} with the MuMax3 package \cite{Mumax2014}.
We use materials parameters of Fe$_{0.7}$Co$_{0.3}$Si \cite{parameter}. By systematically changing the parameter $\alpha$, we observe a sharp transition of the order parameter at the point $\alpha=0$, as shown in Fig.~\ref{fig1}(a). The internal magnetization configuration changes from a FM state to an AFM one, as depicted in the inset. This is a direct evidence of the first-order FM-AFM phase transition. Further, we find that for bilayer ferromagnets with the $\mathcal{P}\mathcal{T}$
symmetry, the first-order phase transition point coincides with
the EP, depending on both the gain-loss parameter and the interlayer coupling constant, as shown in Fig.~\ref{fig1}(b) (see derivations and discussions below).

\begin{figure}[htbp]
	\centering
	\includegraphics[width=0.47\textwidth]{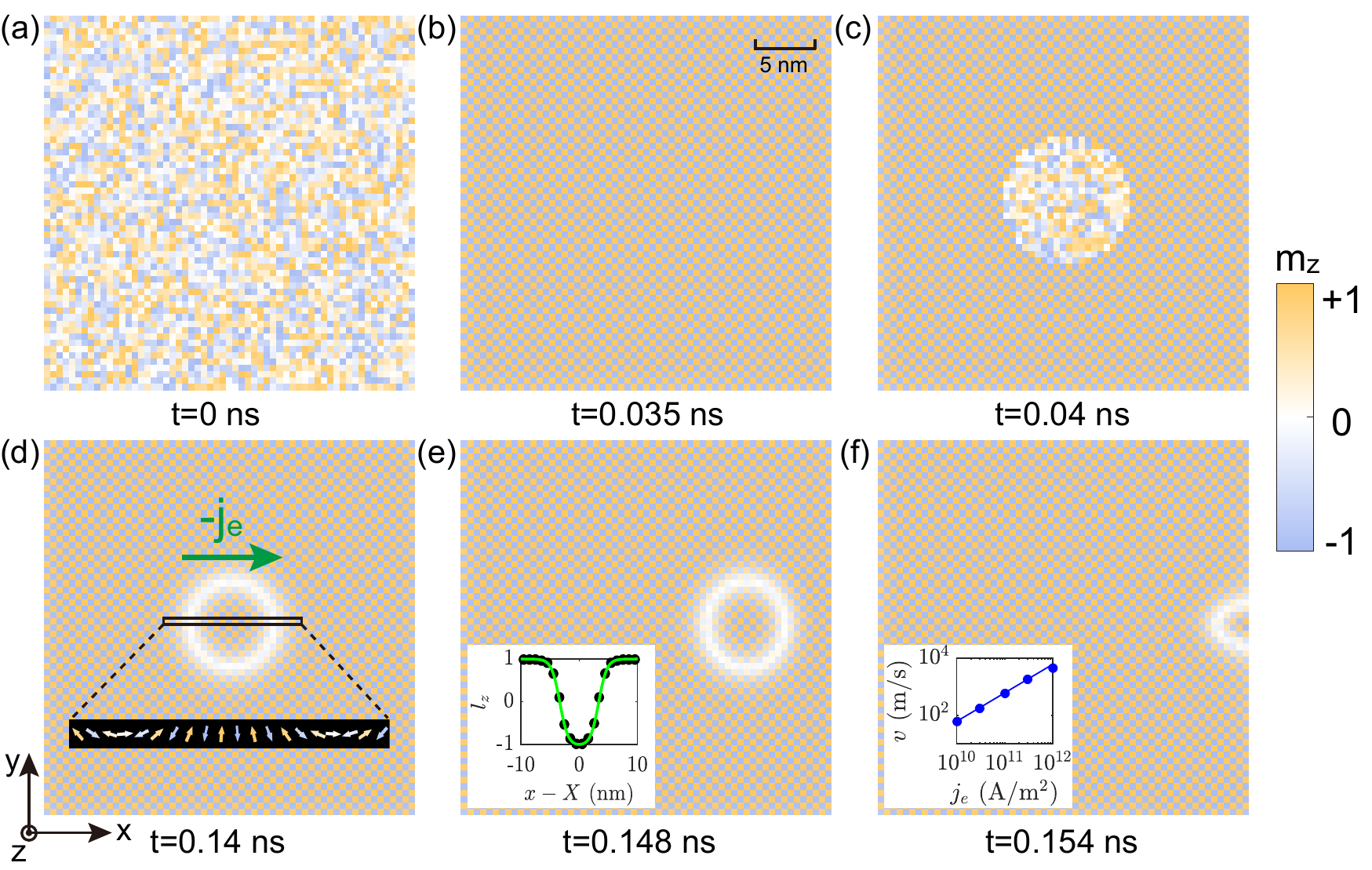}
	\caption{(a) Random spin configuration at $t=0$.
	(b) AFM state evolved at $t=0.035$ ns. (c) Randomizing spins
	inside the circle at $t=0.04$ ns. (d) AFM skyrmion stabilized
	at $t=0.14$ ns. Inset illustrates the magnetization profile
	of the cross section of the AFM skyrmion. (e) Current-driven
	AFM skyrmion motion. Inset displays the spatial distribution of the $z-$component of the N\'{e}el vector
	$\mathbf{l}_i$, with $X$ the position of the skyrmion center. Dots are numerical results and the green curve is a fitting with Eq.~(7) in Ref.~\cite{XSWang2018}.
	(f) AFM skyrmion annihilation at the film boundary. Inset
	shows the current-dependence of the skyrmion velocity when it is far away from the edge. Dots are numerical results and the solid
	line is the analytical formula.
	}
	\label{fig2}
\end{figure}

Now we introduce one compelling application of our findings on skyrmion generations and manipulations.
It is a common wisdom that skyrmions cannot stabilize in a single
easy-plane ferromagnet without applying the external magnetic field
perpendicular to the plane \cite{Vousden2016,Leonov2017,Lin2015}.
We challenge this view by realizing an AFM skyrmion
in a FM thin film with gain. In the simulations, we choose a fixed gain parameter $\alpha=0.01$.

\begin{figure}[htbp]
	\centering
	\includegraphics[width=0.47\textwidth]{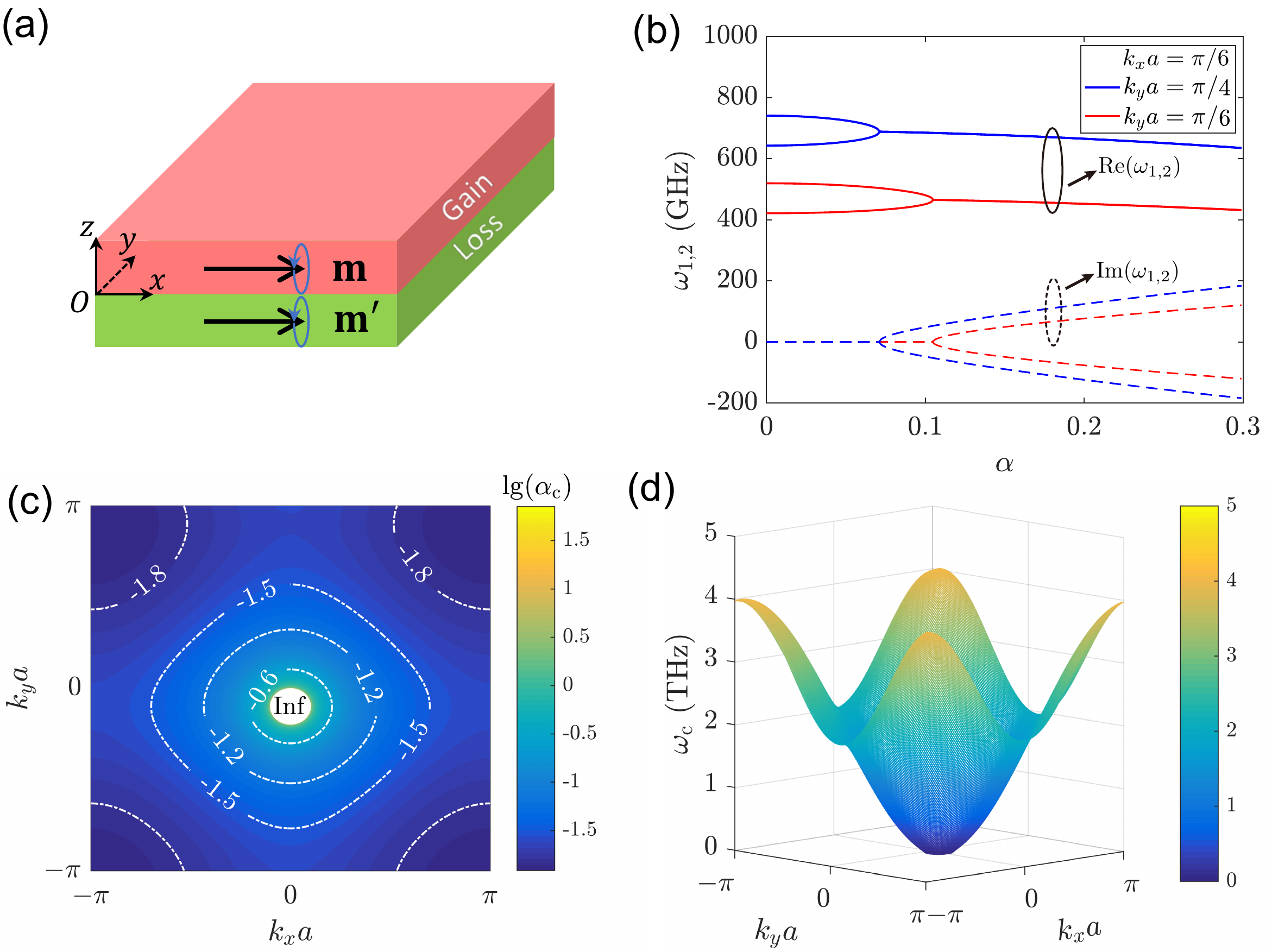}
	\caption{(a) Schematic plot of coupled FM bilayers ($30\times 30\times 1~\hbox{nm}^3$) with balanced gain (red layer) and loss (green layer) with equilibrium magnetizations along $\hat{x}$-direction. (b) Evolution of $\omega_{1,2}$ on $\alpha$ for two representative spin-wave modes $\mathbf{k}=(\frac{\pi}{6a},\frac{\pi}{4a})$ (blue curves) and $(\frac{\pi}{6a},\frac{\pi}{6a})$ (red curves). (c) Contour plot of the mode dependence of $\alpha_{c}$. $\mathcal {P}\mathcal {T}$ symmetry is never broken in the white region labeled as ``Inf" the abbreviation of infinity. (d) Critical frequency $\omega_{c}$ as a function of $\mathbf{k}$. In the calculations, we adopted the materials parameters of Fe$_{0.7}$Co$_{0.3}$Si, and the interlayer coupling constant $\lambda=0.1$. }
	\label{fig3}
\end{figure}

We start with a random initial $(t=0)$
magnetization profile [see Fig.~\ref{fig2}(a)], which
mimics the state of the thermal demagnetization, for
instance. At $t=0.035$ ns, local magnetic moments
quickly evolve to an antiparallelly aligned state,
as shown in Fig.~\ref{fig2}(b). We therefore achieve
an AFM state in a ferromagnet, with the energy
cost $2\times60^{2}J\approx9.4$ eV. The average power is estimated to be as low as $43$ nW, comparable with that to excite spin
waves in FM thin films \cite{BNZhang2018}.
However, the skyrmionic spin texture is yet to emerge.
In Fig.~\ref{fig2}(c) we randomize all spins inside
a circle of radius $5$ nm in the film center, which can
be realized by local heatings or current
pulses \cite{HYYuan2016}. At $t=0.14$ ns, an AFM
skyrmion stabilizes [see
Fig.~\ref{fig2}(d)].
The profile of the staggered magnetization of the AFM skyrmion can be well described by
the formula proposed in Ref.~\cite{XSWang2018},
as shown in the inset of Fig.~\ref{fig2}(e).

To manipulate the AFM skyrmion motion, we apply an in-plane
spin-polarized electric current $\mathbf{j}_{e}=-j_{e}\hat{x}$
with $j_{e}=5.0\times10^{11}$ A m$^{-2}$.
We find that the AFM skyrmion propagates with a large velocity
$3000$ m s$^{-1}$. We note that the skyrmion trajectory is exactly
along the flowing direction of electrons, without suffering
the skyrmion Hall effect [see Fig.~\ref{fig2}(e)]. The high speed of the AFM skyrmion agrees with the formula $\mathbf{v}=(\beta/\alpha) u\hat{x}$ obtained from
the Thiele's equation \cite{Thiele1973,Yang2018}, as shown in the
inset of Fig.~\ref{fig2}(f), where
$u=\mu_{B} j_{e}/[|e|M_{s}(1+\beta^{2})]$ is the drift velocity
of conduction electrons with $\mu_{B}$ the Bohr magneton,
$e$ the electron charge, and $\beta$ the material non-adiabatic
parameter (we set $\beta=0.1$ in the simulations). We do not observe any visible difference with and without dipolar fields. Due to the repulsive force from the boundary, the AFM skyrmion is
slowed down but finally annihilates at the edge since the driving
force from the current overcomes the edge repulsion, as shown in
Fig.~\ref{fig2}(f). All these features of the AFM skyrmion motion
can be well reproduced by simulating \eqref{Recast} instead
of \eqref{MLLG} (not shown).

Compared to their FM counterparts, AFM skyrmions \cite{Bogdanov1989} have some other advantages, such as the elevated mobility \cite{Barker2016,Zhang2016,Jin2016,Velkov2016} and the unusual thermal
properties \cite{Barker2016,Yang2018OE}, among others \cite{Keesman2016,Buhl2017,Gobel2017,Liu2017,Zhao2018}. One recent breakthrough toward this direction is the experimental realization of ferrimagnetic skyrmions in GdFeCo films with inhibited skyrmion Hall effect \cite{Woo2018C,Woo2018E}. Because of its intrinsic difficulties in materials and detections, the AFM skyrmion is yet to be observed in experiments. Our strategy to generate the AFM skyrmion in single-layer ferromagnets thus provides a possible way to overcome the barrier.

We next extend the original idea of Ref. \cite{Lee2015} to two coupled FM films by including finite intralayer exchange couplings, which enables us to investigate the spin-wave (magnon) excitations. Schematic setup is shown in Fig.~\ref{fig3}(a), with $\mathbf{m}$ and $\mathbf{m}'$ representing the spatiotemporal magnetization direction in the layer with gain and the layer with loss, respectively. The equations of motion for the coupled magnetization dynamics read
\begin{equation}\label{CoupledLLG}
  \begin{aligned}
  \frac{d \mathbf{m}_{i}}{d t}=&-\gamma \mathbf{m}_{i}\times \Big[\mathbf{H}_{\text{eff},i}+\lambda (J/\mu_{0}M_{s}a^{3})\mathbf{m}_{i}'\Big]-\alpha\mathbf{m}_{i}\times\frac{d \mathbf{m}_{i}}{d t},\\
  \frac{d \mathbf{m}_{i}'}{d t}=&-\gamma \mathbf{m}_{i}'\times \Big[\mathbf{H}_{\text{eff},i}'+\lambda (J/\mu_{0}M_{s}a^{3})\mathbf{m}_{i}\Big]+\alpha\mathbf{m}_{i}'\times\frac{d \mathbf{m}_{i}'}{d t},
  \end{aligned}
\end{equation}
where $\mathbf{H}_{\text{eff},i}'$ is identical to $\mathbf{H}_{\text{eff},i}$ in Eq. (\ref{MLLG}) by replacing its constituent $\mathbf{m}$ with $\mathbf{m}'$, and $\lambda>0$ is the ratio between the interlayer and the intralayer exchange coupling. Under a combined operation of parity $\mathcal{P}$: $\mathbf{m}_{i}\leftrightarrow\mathbf{m}_{i}'$ and $\mathbf{H}_{\text{eff},i}\leftrightarrow\mathbf{H}_{\text{eff},i}'$ and time reversal $\mathcal{T}$: $t\rightarrow -t$, $\mathbf{m}_{i}\rightarrow -\mathbf{m}_{i}$, $\mathbf{m}_{i}'\rightarrow -\mathbf{m}_{i}'$, $\mathbf{H}_{\text{eff},i}\rightarrow-\mathbf{H}_{\text{eff},i}$, and $\mathbf{H}_{\text{eff},i}'\rightarrow-\mathbf{H}_{\text{eff},i}'$, we find that Eqs. \eqref{CoupledLLG} are invariant and thus respect the $\mathcal {P}\mathcal {T}$ symmetry. To obtain the spin-wave spectrum, we consider a small deviation of both $\mathbf{m}_{i}$ and $\mathbf{m}_{i}'$ from their equilibrium direction $\hat{x}$: $\mathbf{m}_{i}=(1,\delta m_{i,y},\delta m_{i,z})$ and $\mathbf{m}_{i}'=(1,\delta m_{i,y}',\delta m_{i,z}')$ with $|\delta m_{i,y}|+|\delta m_{i,z}|+|\delta m_{i,y}'|+|\delta m_{i,z}'|\ll 1$. The eigensolutions of linearized Eqs. (\ref{CoupledLLG}) have the forms of $\delta m_{i,y}=Ye^{i(\mathbf{k}\cdot \mathbf{r}-\omega t)},\delta m_{i,z}=Ze^{i(\mathbf{k}\cdot \mathbf{r}-\omega t)}$ and $\delta m_{i,y}'=Y'e^{i(\mathbf{k}\cdot \mathbf{r}-\omega t)},\delta m_{i,z}'=Z'e^{i(\mathbf{k}\cdot \mathbf{r}-\omega t)}$ with $\mathbf{r}=(i_{x},i_{y})a$ and $\mathbf{k}=(k_{x},k_{y})$ the wave vector of the spin wave. We thus obtain the equation for the column vector $\Psi(\mathbf{k})=(Y,Z,Y',Z')^{\text{T}}$:
\begin{equation}\label{Eigensolution}
  H(\mathbf{k})\Psi(\mathbf{k})=\omega(\mathbf{k})\Psi(\mathbf{k}),
\end{equation}
where $H$ is a $4\times4$ matrix
\begin{widetext}
\begin{equation}\label{Matrix}
  H(\mathbf{k})=\frac{\gamma}{(1+\alpha^{2})\mu_{0}M_{s}a^{3}}\left(
                                                                 \begin{array}{cccc}
                                                                   \chi_{1}(\mathbf{k})+\alpha [\chi^{*}_{2}(\mathbf{k})-2iK'] & \chi_{2}(\mathbf{k})+\alpha \chi_{1}(\mathbf{k}) & \alpha \chi^{*}_{0}  & \chi_{0} \\
                                                                   -\alpha\chi_{1}(\mathbf{k})+ \chi^{*}_{2}(\mathbf{k})-2iK' & -\alpha\chi_{2}(\mathbf{k})+\chi_{1}(\mathbf{k}) & \chi^{*}_{0} & \alpha\chi^{*}_{0} \\
                                                                   \alpha\chi_{0} & \chi_{0} & \chi_{1}(\mathbf{k})-\alpha [\chi^{*}_{2}(\mathbf{k})-2iK'] & \chi_{2}(\mathbf{k})-\alpha\chi_{1}(\mathbf{k}) \\
                                                                   \chi^{*}_{0} & \alpha\chi_{0} & \alpha \chi_{1}(\mathbf{k})+\chi^{*}_{2}(\mathbf{k})-2iK' & \alpha\chi_{2}(\mathbf{k})+\chi_{1}(\mathbf{k}) \\
                                                                 \end{array}
                                                               \right)
    ,
\end{equation}
\end{widetext}
with $\chi_{0}=i\lambda J,~\chi_{1}(\mathbf{k})=2D\sin k_{y}a,~\chi_{2}(\mathbf{k})=2iJ(\cos k_{x}a+\cos k_{y}a)-i(4+\lambda) J-2iK'$, and $K'=K+\mu_{0}M^{2}_{s}a^{3}/2$ summing up the easy-plane anisotropy and the demagnetizing energy. The solutions of eigenfrequencies come in pairs $\pm\omega$. Two positive solutions, corresponding to counterclockwise magnetization precession around the ground state along $\hat{x}$-direction, are relevant and can be expressed as
\begin{equation}\label{Eigenfrequencies}
  \omega_{1,2}(\mathbf{k})=\lambda+2\zeta(\mathbf{k})\pm \sqrt{\lambda^{2}-4\alpha^{2}\zeta(\mathbf{k})\big[\lambda+\zeta(\mathbf{k})\big]}
\end{equation}
multiplying $\gamma J/\big[(1+\alpha^{2})\mu_{0}M_{s}a^{3}\big]$, with $\zeta(\mathbf{k})=2-\cos k_{x}a-\cos k_{y}a+(D/J)\sin k_{y}a$. In deriving (\ref{Eigenfrequencies}), we have dropped the contribution from $K'$ since we focus on the exchange spin-wave region. For a given $\mathbf{k}$, as the gain and loss parameter $\alpha$ increases, the
two eigenfrequencies approach one another, and at some critical value $\alpha=\alpha_{\text{c}}$ they coalesce at the EP and bifurcate into
the complex plane [see Fig. \ref{fig3}(b)]. At the EP, the two normal modes coalesce as well. The domain with real eigenfrequencies is termed the \emph{exact phase}, otherwise it is called the \emph{broken phase}. From Eq. (\ref{Eigenfrequencies}) one can obtain both the critical gain-loss parameter and the critical frequency. Nevertheless, we point out a special region $-\lambda\leqslant\zeta(\mathbf{k})\leqslant0$ in which the $\mathcal {P}\mathcal {T}$ symmetry is \emph{never} broken without considering the nonlinear effect of the LLG equations (\ref{CoupledLLG}). This fact is in contrast to conventional $\mathcal {P}\mathcal {T}$ symmetric systems suffering symmetry breaking when the strength of the gain-loss term exceeds a certain critical value \cite{Lee2015}. Of course, the nonlinear magnon-magnon interaction complicates this picture and will generate a level broadening of spin-wave eigenmodes \cite{BNZhang2018}. For $\zeta(\mathbf{k})$ outside $[-\lambda,0]$, the two critical parameters are given by
\begin{equation}\label{Critical}
  \alpha_{\text{c}}(\mathbf{k})=\frac{\lambda}{2\sqrt{\zeta(\mathbf{k})\big[\lambda+\zeta(\mathbf{k})\big]}},~~
  \omega_{\text{c}}(\mathbf{k})=\frac{\gamma J}{\mu_{0}M_{s}a^{3}}\frac{\lambda+2\zeta(\mathbf{k})}{1+\alpha_{\text{c}}^{2}(\mathbf{k})},
\end{equation}
both of which are mode-dependent. Figures \ref{fig3}(c) and (d) show the distribution of $\alpha_{\text{c}}$ and $\omega_{\text{c}}$ over the first Brillouin zone, respectively. The center of the white region in Fig. \ref{fig3}(c) does not coincide with the origin, with a downward shift $\arctan(D/J)$ caused by the DMI.

\begin{figure}[htbp]
	\centering
	\includegraphics[width=0.5\textwidth]{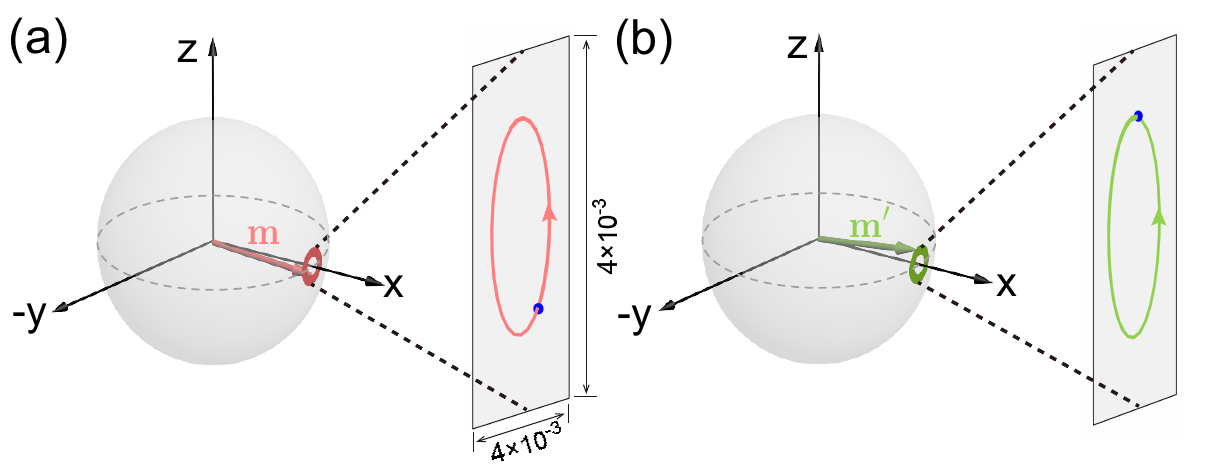}
	\caption{Trajectory of the steady-state magnetizations at the site $(30a,30a)$ in both the ``gain" layer (a) and the lossy layer (b) in the exact phase, with zoomed in details shown in the right side. The blue dot indicates the instantaneous phase of the spin. We set pinned boundary conditions and $\alpha=0.01$ with rest parameters the same as those used in Fig. \ref{fig3}.}
	\label{fig4}
\end{figure}

In the exact phase $\alpha<\min_{\forall\mathbf{k}}\alpha_{\text{c}}(\mathbf{k})=\lambda/\big[2\sqrt{\zeta_{0}(\lambda+\zeta_{0})}\big]$ with $\zeta_{0}=3+\sqrt{1+(D/J)^{2}}$, predictions from the linear spin-wave theory compare well with the full simulation of Eqs. (\ref{CoupledLLG}) that the steady-state magnetizations in both layers oscillate around the initial misalignment from the $\hat{x}$ axis without being attenuated or amplified (see Fig. \ref{fig4}). Since both layers are in the FM state in the exact phase, we only observe the counterclockwise spin-wave modes.

\begin{figure}[htbp]
	\centering
	\includegraphics[width=0.47\textwidth]{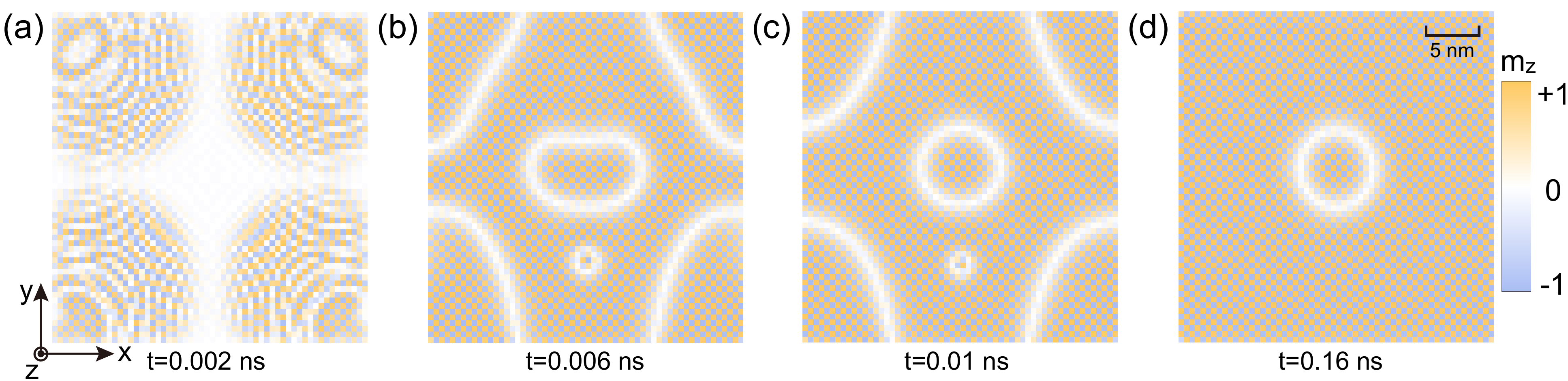}
	\caption{Time evolution of magnetizations in the ``gain" layer under the broken phase, at (a) $t=0.002$ ns, (b) $t=0.006$ ns, (c) $t=0.01$ ns, and (d) $t=0.16$ ns. $\alpha=0.3$ in the simulation.}
	\label{fig5}
\end{figure}

In the broken phase $\alpha>\lambda/\big[2\sqrt{\zeta_{0}(\lambda+\zeta_{0})}\big]\approx0.012$ for $\lambda=0.1$, the linear theory indicates an exponential growth of the spin-wave amplitude, which is associated with the case that the eigenfrequencies (\ref{Eigenfrequencies}) have an imaginary part. The induced instability can drive the spin away from its equilibrium direction. The situation in the critical phase $\alpha=\lambda/\big[2\sqrt{\zeta_{0}(\lambda+\zeta_{0})}\big]$ is similar: The linear spin-wave theory introduces a linear instead of exponential growth of the wave amplitude, which is the consequence of the EP degeneracy. The lossy layer thus preserves the in-plane FM state to some extent (not shown). However, in the gain layer, it is interesting that the original in-plane magnetizations along the $\hat{x}$-direction evolve to be perpendicular to the plane (but with a negligibly small global canting of an angle less than 2.5$^{\circ}$ with respect to the normal of the lossy layer), and finally form an AFM skyrmion, as showin in Figs.~\ref{fig5}(a)-(d). The EP thus exactly coincides with the FM-AFM phase transition point in the gain layer, which has been verified by micromagnetic simulations, see Fig.~\ref{fig1}(b).

Negative damping is essential to realize our proposal. Its real world implementation methods are multiform besides the two approaches introduced above. A recent experiment reported the electricfield-induced negative negative magnetic damping in FM$|$FE (ferroelectric) heterostructures \cite{Jia2015}. In Ref. \cite{Wegrowe2007}, Wegrowe \emph{et al}. thoroughly analyzed the spin transfer in an open FM layer, and found that the negative damping appears naturally for describing the exchange of spins between the magnetic system and the environment \cite{Li2003,Ando2008,Duan2014}.

In summary, we uncovered a mapping between a ferromagnet with gain and an antiferromagnet with an equal amount of loss. A novel first-order FM to AFM phase transition, or vice versa, was predicted by tuning the gain-loss parameter. In a chiral easy-plane ferromagnet in the presence of gain, we showed the emergence of a stabilized AFM skyrmion without applying any external field. In 1D and 2D non-chiral ``gain" ferromagnets, we envision the formation of AFM domain walls \cite{Li2014} and vortices \cite{Wu2011}, respectively. We also studied the spin-wave spectrum in FM bilayers with balanced gain and loss. We predicted a spectral region in the first Brillouin zone, in which the $\mathcal {P}\mathcal {T}$ symmetry is never broken in the framework of linear spin-wave theory. We found that the emerging antiferromagnetism and the AFM skyrmion appear in the ``gain" layer only in the cases of broken $\mathcal {P}\mathcal {T}$ symmetry phase. The results presented here open a new way to create and manipulate AFM solitons in simple ferromagnets through the first-order FM-AFM phase transition, and build a novel bridge connecting the $\mathcal {P}\mathcal {T}$ symmetry to magnonics and skyrmionics.

\begin{acknowledgments}
This work is funded by the National
Natural Science Foundation of China (Grants No. 11704060 and 11604041), the
National Key Research Development Program under Contract No. 2016YFA0300801,
and the National Thousand-Young-Talent Program of China.
C.W. acknowledges the financial support from the China Postdoctoral
Science Foundation (Grants No. 2017M610595 and 2017T100684) and the National
Natural Science Foundation of China under Grant No. 11704061.

H.H.Y. and C.W. contributed equally to this work.
\end{acknowledgments}

\end{document}